# A Review of Artificial Intelligence Technologies for Early Prediction of Alzheimer's Disease

Kuo Yang,     Emad A. Mohammed

*Abstract*— **Alzheimer's Disease (AD) is a severe brain disorder, destroying memories and brain functions. AD causes chronically, progressively, and irreversibly cognitive declination and brain damages. The reliable and effective evaluation of early dementia has become essential research with medical imaging technologies and computer-aided algorithms. This trend has moved to modern Artificial Intelligence (AI) technologies motivated by deep-learning success in image classification and natural language processing. The purpose of this review is to provide an overview of the latest research involving deep-learning algorithms in evaluating the process of dementia, diagnosing the early stage of AD, and discussing an outlook for this research. This review introduces various applications of modern AI algorithms in AD diagnosis, including Convolutional Neural Network (CNN), Recurrent Neural Network (RNN), Automatic Image Segmentation, Autoencoder, Graph CNN (GCN), Ensemble Learning, and Transfer Learning. The advantages and disadvantages of the proposed methods and their performance are discussed. The conclusion section summarizes the primary contributions and medical imaging preprocessing techniques applied in the reviewed research. Finally, we discuss the limitations and future outlooks.**

## I. INTRODUCTION

Alzheimer's Disease (AD) has been one of the most common forms of dementia, which could cause cognitive damages, memory disorders, memory loss and difficulties in decision-making, verbal communication, concentration, thinking and judgement. The AD symptoms progress slowly over time to undermine patients' living ability to perform daily tasks. So far, there is no complete cure for AD, and all treatments are to slow AD-related symptoms worsening based on patients' symptoms. However, these treatments add an endless financial burden on patients, their families, and the government health care system. Thus, it is essential to achieve a reliable and efficient method to detect AD as early as possible. Patients could then have immediate treatments to slow down the symptom progress early, avoiding the worst dementia effects.

According to the latest survey report, AD affected more than 30 million individuals in 2015, and this number could exceed 114 million by 2050 [1]. Dementia caused by AD can result in a loss of tissues across all brain regions, leading to significant damage to the neural system, breaking down the neural functions to negatively affect the patient's cognitive ability.

The tissue loss starts in Grey Matter (GM), and this loss spreads out to White Matter (WM), Corpus Callosum (CC) and Hippocampus (HC) [2] gradually. Thus, the early stage of AD can be diagnosed and detected by analyzing the brain's specific structures' variations. Individuals with Mild Cognitive Impairment (MCI) have the highest risks of progressing into the finally irreversible brain disorder stage [3]. For the early prediction of AD, MCI could subject to stable MCI (sMCI) and progressive MCI (pMCI). The complete progress of AD-related dementia starts from Normal Cognitive (NC), sMCI to pMCI and finally AD.

Magnetic Resonance Imaging (MRI), including structural MRI (sMRI) and functional MRI (fMRI), are the primary medical imaging tools to help understand and evaluate the anatomical changes of sensitive regions related to AD [4]. Therefore, MRI is the most significant source for medical professionals to diagnose AD in hospitals, and it is also an essential data source in AD detection research. The longitudinal data based on subjects' cognitive measures (NC, MCI and AD) contribute to diagnosing AD in the early stage [5]. This type of data typically has a time-length of 6 to 12 months. It is made of various medical examinations and measures, such as Assessment Scale-Cognitive Subscale (ADAS-Cog 13), Rey Auditory Verbal Learning Test (RAVLT), Functional Assessment Questionnaire (FAQ) and Mini-Mental State Examination (MMSE). In addition to this, as an essential part of clinical care, medical history becomes a rich source for early detection of AD in literature. This type of data includes Electronic Health Records (EHRs) and clinical recordings [6]. Multi-modality images help improving AD diagnosis, including both MRI and Positron Emission Tomography (PET) [7]. As discussed above, MRI can provide sufficient information related to anatomical structures across all brain regions. PET scan can indicate the brain's metabolic activities, providing valuable information related to AD-related dementia prediction.

AD prediction is to classify different stages of dementia progress, including NC, MCI (sMCI and pMCI) and AD, so the early prediction becomes a pattern classification problem [3]. Many modelling techniques, such as statistical models and machine learning algorithms, have been explored with hand-crafted features to improve the model's capability to understand more extensive health conditions. However, recent research focuses on the deep-learning algorithm for the AD early diagnosis due to its excellent automatic feature

K. Yang is with the School of Engineering, Lakehead University, Thunder Bay, ON, P7B5E1, Canada (e-mail: kyang3@lakeheadu.ca)

E. A. Mohammed with the School of Engineering, Lakehead University, Thunder Bay, ON, P7B5E1, Canada (e-mail: emohamme@lakeheadu.ca)

extraction and representation of the image or text data, as shown in Table I. Previous research mostly focuses on applying deep-learning algorithms such as the CNN model to predict AD fundamentally. Simultaneously, the latest studies concentrate on applying more complicated multi-channel CNN models, pre-trained CNN architectures, CNN / RNN autoencoders, or combining both CNN and RNN models to improve the model performance. Therefore, with this fast growth in deep-learning technology, recent researches focus on predicting AD's early stages more precisely.

As the AI field (especially deep-learning) is a fast-growing field, we focus the review scope on the latest published papers to create a comprehensive summary of the most recent effort to predict AD at the early stages. We focus on applying AI technologies to predict AD (rather than explain the AI technologies' details) to filter this review's initial papers, as illustrated in Table I.

TABLE I
CATEGORIES OF ARTIFICIAL INTELLIGENCE (AI) TECHNIQUES IN AD DETECTION RESEARCHES

| AI Techniques | References |
|---|---|
| CNN | [3], [8], [9], [10], [11], [12], [13], [14], [15], [16], [17], [18], [26] |
| RNN | [5], [6], [19], [20] |
| GCN | [21], [22] |
| Ensemble Learning | [4], [7], [23], [24] |
| Transfer Learning | [27], [28], [29], [30] |
| Others | [2], [25], [31], [32], [33] |

The primary deep-learning technology applied Convolutional Neural Network (CNN) to classify AD vs NC vs MCI. CNN based autoencoder is also involved in generating the low-dimension representation of MRIs or PETs [16], and the classification or regression is realized via traditional machine learning algorithms such as Support Vector Machine (SVM) [12]. As one application of CNN, Semantic Image Segmentation (SIS) can potentially highlight particular brain regions on MRIs related to AD progression, and it also gives a classification of the subject data. Recurrent Neural Network (RNN) is employed to extract the text data patterns or compress the data into a deep representation to take advantage of patients' medical history. Both CNN and RNN needs a large number of training data and optimized structure to achieve reliable performance. Transfer Learning (TL) can take advantage of the pre-trained weights and model architecture into AD detection with new data [31]. Graph data can combine both medical imaging data and patients' information into a single network. Each subject is then represented as a node and can be analyzed for AD status via Graph Convolutional Neural Network (GCN) [21]. Ensembling Learning Method is to combine all classifiers' outputs using aggregation method such as majority voting, weighted majority voting, to minimize the prediction variance.

The remainder of this review is organized as follows. Section II discusses the AI technologies applied in early AD diagnosis research, compares research contributions, discusses the methods' advantages and disadvantages. Finally, in section III, an in-depth conclusion of this literature review is illustrated with our perspective to enhance AD's early detection using a newly introduced deep-learning approach.

## II. LITERATURE REVIEWS

### A. Convolutional Neural Network (CNN) for AD prediction

Habes, et al. proposed a new framework to use CNN to classify AD and Normal Cognitive (NC). The proposed work applied feature maps of CNN to build a time-to-event prognostic model with extra clinical variables related to AD dementia, and this model can predict the patients' progress to AD [3]. The dataset used in this study was MRI images acquired from ADNI[1] and AIBL[2]. The hippocampus region was extracted from 3D MRIs via local label learning algorithm, and the 3D volume was sliced to form the training dataset. In the proposed method, the CNN model took both left and right extracted hippocampal images as inputs for the label classification. The trained CNN feature maps are combined with the related clinical variables to build the LASSO regularized Cox regression model [34]. The proposed method included both CNN and a regression model, and the regression model was trained with the feature maps from CNN. However, two models were trained separately instead of jointly, and this made it difficult to integrate both regressor and classifier as a complete framework.

Z. Cui et al. proposed an enhanced inception network to predict AD's early stage based on brain MRI [8]. Image preprocessing included histogram equalization and multi-threshold segmentation [35] to highlight AD's sensitive brain regions. The enhanced inception model involved one additional branch composed of two convolution layers (3x3 kernel) with sigmoid activation, and this branch's output was multiplied with the base inception output. This additional branch increased the receptive field of images. It generated the corresponding attention heat map with all values within range [0,1], so this heat map highlights the attention region of MRIs, such as the Hippocampus Region. However, this enhanced model lacked a specific mathematical explanation of how the model highlights certain regions of MRI.

S. Basaia et al. built a deep-learning model predicting the individual diagnosis of AD and Mild Cognitive Impairment (MCI) based on a single cross-sectional brain structural MRI scan [9]. This work applied data augmentation techniques (e.g., rotation, translation of the images) to fix the overfitting issue and achieved an excellent result.

M. Amin-Naji et al. applied the Siamese Convolutional Neural Network (SCNN) to classify AD and NC patients based on MRI images [10]. The proposed SCNN had three branches of ResNet-34 to generate three vectors, and these vectors were the output feature maps that were used for AD early prediction. This method selected the MRI images randomly from either AD or NC dataset to feed SCNN on the anchor and positive branches for two vectors (anchor vector, positive vector). These two vectors were used to calculate one positive pairwise distance. In contrast, the negative branch's images have labels against positive and anchor branches to generate the negative vector. However, the negative pairwise distance was calculated with both the negative vector and the anchor vector. For example, the negative branch picks NC subjects' images when both positive and anchor branches pick AD subjects' images. Both positive and negative pairwise distances were used in the loss function, as depicted in equation 1 [10].

$$L(anch, pos, neg) = \max (||f(anch) - f(pos)||^2 - ||f(anch) - f(neg)||^2 + \alpha) \qquad (1)$$

The value α is a margin value. The model training's objective is to maximize the difference between pairwise distances, which means the distance between the anchor and positive images decreases with the distance between the anchor and negative images increasing. The proposed work contributed to apply the unsupervised learning method to classify AD and NC subjects. However, the limitation was that the SCNN was a binary classifier, and it was restricted to predict AD vs NC vs MRI.

M. Liu et al. proposed a model combining a classifier and a regressor to predict AD [11]. One part of the training data was the patches of multiple anatomical landmarks, and the other part was hand-crafted features, including four clinical scores and two demographic factors. The proposed model started with a 2-channel CNN, which was trained with landmark patches. In the final layer, both CNN feature maps and the hand-crafted features were concatenated to form a multi-classifier (labels: AD pMCI, sMCI and NC) and a regressor to predict clinical scores. The proposed work applied CNN for both classification and regression.

Based on [11], M. Liu et al. proposed a method to predict clinical scores using MRI data and the same CNN architecture [13]. The data was composed of landmark patches extracted from the MRI. However, in this study, the clinical score was incomplete in the dataset, so the proposed work's contribution was to apply deep-learning algorithms to predict the incomplete clinical score. However, the missed clinical score update was based on the complete clinical score, so the regression accuracy was limited.

F. J. Martinez-Murcia et al. proposed a new method to analyze the AD status via Deep Convolutional Autoencoders (DCA) [12]. The MRI was preprocessed to segment Grey Matter (GM) and White Matter (WM) regions. DCA was composed of an encoder and a decoder. The encoder compresses MRI slices into low-dimensional representations via the down-sampling CNN, while the decoder was reconstructed via the up-sampling CNN. The training process objective is to minimize the Euclidean distance between the original and reconstructed MRI, improving compressed images' quality. The new representation of AD and NC subjects were classified via SVM and Neural Network (NN). The major limitation of DCA was that the model minimized the loss between input and output images and failed to extract the statistical patterns of different MRI categories in the image representation.

X. Hong et al. proposed a new architecture of CNNs to classify AD and Healthy Cognitive (HC) [16]. The proposed work combined both 2D and 3D CNNs into 2 CNN models, and the models shared the feature maps in the middle of the computing process. Then the final layer averaged both models' output feature maps to predict results. The proposed method applied both 2D and 3D MRI in the CNN model, but this framework relied on a balanced dataset and sufficient MRI images.

M. Kavitha et al. investigated a modified U-net-like architecture to perform both segmentation and classification on Positron Emission Tomography (PET) [26]. The proposed mode constructed the bottleneck architecture of the CNN model to extract feature maps of PET images. The model's final layer was a multi-class logistic regression classifier to predict AD, NC, and MCI.

Recent research studied the impacts of different optimizers, overfitting techniques, and various CNN model architectures to classify AD stages. A. M. Taqi et al. explored data augmentation and several optimizers to classify AD and Normal Cognitive (NC) [14]. L. Yue et al. proposed a method to use a custom CNN to predict AD [15]. U. Senanayake et al. also proposed a similar CNN architecture to predict the early stages of AD with dense block and residual block nets [36]. Daniel et al. investigated a 2-channel 1D Convolutional Neural Network to predict AD via hand-featured data [17]. A. Fedorov et al. applied the Deep InfoMax (DIM) algorithm to generate deep representations of MRIs [18]. DIM is unsupervised learning of representations by maximizing mutual information between input and output of a deep neural network encoder [37]. These researches proved that data augmentation could decrease overfitting effectively. Adam optimizer and RMSProp optimizer achieved the best results. However, the proposed methods only focused on classifying AD and NC, not investigating the model performance in early-stage predictions such as MCI.

## B. Graph Convolutional Network (GCN) for AD Prediction

X. Zhao et al. introduced a method based on GCN to predict Mild Cognitive Impairment (MCI) based on the resting-state functional magnetic resonance imaging (rs-fMRI) [21]. The dataset of rs-fMRI in this study was from ANDI, including Late Mild Cognitive Impairment (LMCI), Early Mild Cognitive Impairment (EMCI) and Normal Control (NC). The image preprocessing was processed by the GRETNA[3] and DARTEL[4], including converting DICOM to NIFTI, time slicing, realignment. The proposed method started with extracting the functional connectivity coefficient matrix from Brain Functional Connectivity (BFC5) from rs-MRI. Each matrix combined with subjects' gender, scan device information and labels to generate the subject vectors. Each vector was considered a node, and all connections were generated based on the similarity among these nodes. Finally, the complete graph data was composed of nodes and node edges.

J. Guo et al. applied GCN to predict the early stage of AD on Positron Emission Tomography (PET) [22]. Image preprocessing was used to generate Region Of Interests (ROI)

---

on PETs. Furthermore, each PET was converted into graph data by calculating the correlation of the data. Both [21] and [22] investigates how the graph data was applied for AD prediction. Both sMRI and PET were used to generate graph data, and the significant difference was that the node was classified in one complete graph in [21], while each graph was classified in [22]. However, these researches relied on third-party medical imaging packages to generate and integrate data with the data processing framework.

## C. Ensemble Learning for AD Prediction

F. Li et al. applied the ensemble learning method to predict the early stage of AD [4]. The proposed method uniformly partitioned 3D-MRI images into 27 regions, and 3D image patches of each region were extracted to form the training dataset. The Principle Component Analysis (PCA) was used to reduce the dimension of each patch data vector. Then, the extracted patches of each region were clustered by the K-means clustering algorithm. In each cluster, one Dense-Net CNN was connected to perform classification. After that, all the pre-trained Dense-Net CNNs were connected to a dense layer with SoftMax to perform classification tasks. This research used a fully dense layer to ensemble the pre-trained models' outputs matching the target label. However, this ensembling was separated from the framework, and it cannot adapt to dataset changes or updates. Also, each cluster of data involved a DenseNet training process, which caused high computation costs.

M. Naji et al. proposed a hybrid ensemble learning method to predict AD [23] and MRI, including AD, MCI and CN. The hybrid method involved an image optimization selection process and one CNN model. The image selection was optimized via the image entropy algorithm [39] to filter out the most non-informative MRI images. The proposed CNN model had three channels accepting MRI images on the axial, coronal and sagittal planes individually. The three channels' outputs were embedded via the soft voting technique to give the final prediction. The proposed method increases the variety of ensembling learning model by applying MRI on three planes. However, the voting technique cannot adjust weights on the different channels' predictions in the final ensembling.

J. Choi et al. proposed a new ensemble generalization loss function to predict AD based on MRI [24]. The proposed method used MRI images of three planes for training purposes. The model had three channels, and each channel was composed of VGG16, GoogleNet and AlexNet architectures. Then MRI images on the three planes were fed into three independent CNN channels. The proposed loss function is as depicted in equation 2.

$$f(w) = -log \sum_{k=1}^{M} w_k P_k(C_i|x_i^P) -$$
$$\gamma \sum_{k=1}^{M} w_k \sqrt{\frac{1}{M} (\sum_{k \neq n} (P_k(C_i|x_i^P) - P_n(C_i|x_i^P))^2)} \quad (2)$$

where $w_k$ is the weights on each branch output, k is the number of output branches. $x_i^P$ is the image data at P planes (Axial, Sagittal and Coronal planes) and $P_k(C_i|x_i^P)$ is predicted probability of label $C_i$ given an image input $x_i^P$. The proposed loss function has two terms. The first term is to calculate the loss of all branch outputs, and the second term is to get an average of all prediction distances in the output. Parameter $\gamma$ is to control the contribution of two terms in the loss function. If $\gamma$ is 0, the model only considers the weights of each branch in the final ensemble. However, if $\gamma$ rises, weights are increased by considering the distance of all branch predictions. The proposed loss function combines both model training and model ensembling as a whole framework. Also, updates of ensembling weights have pre-trained models' outputs adapted to target labels with various effects. However, the loss function does not involve the target label to calculate the loss, and the model is trained in an unsupervised way, so the training process has no consideration of image labels, which might negatively affect model performance.

X. Fang et al. ensembled three CNNs with the AdaBoost algorithm [40] with multi-modality images for AD classification [7]. The multi-modality images in this research included MRI and PET images. The proposed work had both PET and MRI images fed into 2- channel CNN and each channel was composed of GoogleNet, ResNet and DenseNet. Finally, the classifier predictions were fed into the AdaBoost model to improve accuracy. Models trained with MRIs and PETs were ensembled via the AdaBoost technique. However, AdaBoost focuses on misclassification data, so this could bias model learning on the data noise.

## D. Recurrent Neural Network (RNN) for AD Prediction

H. Li et al. proposed a prognostic model to predict the early stage between MCI and AD [5]. The image dataset was longitudinally measured MRI, including sMCI6 and pMCI7. The proposed method applied Long Short-Term Memory (LSTM) autoencoder [41] to generate deep MRI image representations. The representation combined with hand-crafted-features such as hippocampal volume measures and demographics generated data for the regression model. Finally, the Cog regression model was built to predict the clinical score for AD prediction. S. Fouladvand et al. also applied LSTM autoencoder to generate patient information representation with MRI [42]. In contrast, the proposed method applied the K-means clustering t-distribution Stochastic Neighborhood Embedding algorithm (t-SNE) to classify the data representation. The proposed work in both [41] and [42] heavily relied on time-series related to MRI measurement, but it is challenging to collect quality data with full longitudinal MRI for each subject. Therefore, a lack of data or incomplete data could be a significant limitation for the model performance.

G. Lee et al. investigated the Multi-modal Recurrent Neural Network (MRNN) to analyze the conversion from MCI to AD [19]. The MRNN refers to RNN trained by multi-model data, including demographic information, neural imaging phenotypes, cognitive performance, and Cerebro Spinal Fluid (CSF) measurements in time series. Each type of data was fed into one Gated Recurrent Unit (GRU) in a time sequence for training, and all outputs were concatenated for final prediction. Multi-modal data increased the dataset's variety, enabling MRNN to make the prediction based on complete subject information. However, the training data for MRNN was hand-crafted features instead of automated feature extraction from the MRI images, which required highly professional views in data preprocessing. Also, the lack of complete time-series data was an existing limitation, as in [41] and [42]. Due to the discontinuous longitudinal dataset, C. K. Fisher et al. applied RNN to fill up the incomplete patients' clinical notes or information [43].

---

[6] sMCI: stable MCI who remained as MCI at the last visit
[7] pMCI: progressive MCI who converted to AD before the last visit

The research involved in this section employed RNN to perform classification or generate data representation and the training data was primarily composed of longitudinal data. However, the research focused on combining various essential aspects of AD patients' information in both texts and images but not to generate data embeddings in depth. Therefore, weak embeddings of text data or images could obstruct the interpretability of the RNN model immensely.

### E. CNN Segmentation for AD Prediction

D. Chitradevi et al. analyzed brain sub-regions via segmentation techniques to predict the early stage AD [2], and the MRI dataset included AD and Normal Cognitive (NC) patients. MRIs were processed by skull stripping and histogram equalization to increase the image quality. The multilevel thresholding algorithm [44] was applied to segment the specific regions, including White Matter (WM), Corpus Callosum (CC), Grey Matter (GM) and Hippocampus Regions (HR) on MRI. Then the segmented images were fed into CNN for the AD prediction. The proposed framework involved multilevel thresholding in highlighting AD-related brain regions on MRI boosting the CNN performance. However, this automatic thresholding algorithm required a high contrast between background and foreground on MRI slices. Also, the applied segmentation algorithm was isolated from CNN model training.

E. Lee et al. introduced a method that integrates voxel-based, region-based and patch-based approaches into a unified framework with a deep-learning model to predict AD [25]. The proposed work segmented three tissues on each MRI, including Gray Matter (GM), White Matter (WM) and Cerebro Spinal Fluid (CSF). Each region was interpreted via a deep-learning network for a prediction score related to an AD status, and all scores were averaged to determine the final output eventually.

This section's research enabled AI models to segment ROI on the input images for both classification and regression. However, the segmentation relied on the hypothesis of target regions in the image preprocessing, and the segmented data has to be inspected by specific professionals to assert the region's accuracy.

### F. Transfer Learning for AD Prediction

M. Han et al. applied transfer learning to classify AD vs NC with VGG19 and Inception V4 [30]. N. M. Khan et al. also investigated AD's detection based on MRIs via transfer learning techniques [27]. The proposed work optimized the image selection via the image entropy algorithm [39]. Also, different configurations of VGG pre-trained models were tested to overcome over-fitting issues.

A. Ebrahimi-Ghahnavieh et al. proposed a new structure combining CNN and Long Short-Term Memory (LSTM) to predict AD via transfer learning [28], and the dataset was the sliced MRIs from three different planes. Pre-trained models were constituted with GoogleNet, AlexNet, VGG-Net, Squeeze-Net, Resnet and Inception V3. The proposed study showed that the Squeeze-Net with LSTM acheived the best performance.

K. Aderghal et al. investigated a cross-modal data preprocessing from Structural MRI (MRI) to diffusion tensor imaging modality [29] for the transfer learning model. The image preprocessing included segmenting HR, GM and WM. The diffusion tensor images were generated to couple with MRIs for model training. The proposed method increased the variety of data for model training, but it applied new data to the previous research model structure instead of classic CNN architecture.

Transfer learning increased both model training speed and model performance in AD vs MRI vs NC classifications. However, the model's accuracy profoundly depends on medical imaging preprocessing to highlight the AD-related regions on either MRIs or PETs. Also, the amount of data could cause overfitting and weaken the pre-trained model.

### G. Other Techniques for AD Prediction

J. Albright et al. proposed an all-pairs technique to process hand-crafted features related to the AD status [31]. These features were used to train applied Support Vector Machine (SVM), Logistic Regression Classifier (LRC), Random Forests (RF), Neural Network (NN) classifying AD vs Normal Cognitive (NC). A. Abrol et al. also trained SVM to classify AD vs NC with hand-crafted time-series data extracted from MRIs [32]. X. Hao et al. applied Random Forests (RF) to optimize feature selection for model training in the AD prediction[33].

## III. DISCUSSION AND CONCLUSION

### A. Artificial Intelligence (AI) Technologies for AD Prediction

All reviewed papers are the latest research related to applying deep-learning technologies in the AD prediction area. Based on significant contributions of these researches, there are six AI technologies explored, including Convolutional Neural Network (CNN), Graph Convolutional Neural Network (GCN), Ensembling Learning Methods (ELM), Recurrent Neural Network (RNN), Image Segmentation (IS), Transfer Learning Methods (TLM).

Recent research employed various CNN architectures, such as Inception, ResNet, DenseNet, for better model performance. This research included testing the effect of data augmentation, optimizers under various CNN architectures related to the AD prediction accuracy [9] [17]. These researches used CNN to extract essential features of MRI or acquire the dense representation of MRI to build a regression model for AD score prediction or to train a different classifier [3] [12] [15]. Due to CNN's excellent performance on image classification, more researches used several data modalities

TABLE II
CATEGORIES OF AI TECHNIQUES FOR AD PREDICTION

| AIT | Task | Ref. | Major Contributions | Data Types |
|---|---|---|---|---|
| CNN | Reg. | [3] | LASSO regularized Cox regression model [2] with CNN output features | MCI & NC. (MRI) and Clinical Variables |
| | Clas. | [8] | Improved Inception V3 model | AD & NC, MCI & NC, AD & NC & MCI (MRI) |
| | Clas. | [9] | Data augmentation | AD & NC (MRI) |
| | Clas. | [10] | 3-channel SCNN | AD & NC (MRI) |
| | C&R | [11] | 2-channel CNN for both MRI classification and clinical score regression | pMCI & sMCI & AD & NC (MRI) and Clinical Variables |
| | Clas. | [25] | CNN Autoencoder for MRI compression | AD & NC, sMCI & pMCI (MRI) |
| | Reg. | [13] | Semi-supervised learning with 2-channel ResNet CNN for clinical score regression | AD & MCI & NC (MRI) |
| | Clas. | [14] | Data augmentation and different optimizer comparing | AD & NC (MRI) |
| | Clas. | [15] | Feature extraction with deep CNN | AD & MCI & NC (MRI) |
| | Clas. | [16] | Jointly train CNN model with 2D & 3D MRI | AD & MCI & NC (MRI) |
| | Clas. | [18] | Deep InfoMax to generate the deep representation of MRI | AD & MCI & NC (MRI) |
| | Clas. | [17] | Interpret text-based data via 1-D CNN | AD & NC (MRI) |
| GCN | Clas. | [21] | Graph Cheb-Chevy Neural Network | AD & NC (MRI) |
| | Clas. | [22] | Graph Cheb-Chevy Neural Network | AD & NC (MRI) |
| Ensemble Learning | Clas. | [4] | Ensemble multiple DenseNets trained by various regions of MRI | AD & NC (MRI) |
| | Clas. | [23] | Image entropy [16] and ensemble multiple CNNs trained by MRI on three planes | AD & NC & MCI (MRI) |
| | Clas. | [24] | Deep ensemble generalization loss function | AD & NC & MCI (MRI) |
| | Clas. | [26] | Ensemble CNN prediction with AdaBoost | AD & NC (MRI) |
| RNN | Reg. | [5] | LSTM autoencoder and Cox regression model | Longitudinal Clinical Variables |
| | Clas. | [19] | Ensemble multi-GRUs for prediction | Modality Clinical Variables |
| | Clas. | [6] | LSTM autoencoder to generate the representation of text-based data | AD & NC (MRI) |
| | Clas. | [20] | Fill up incomplete values of text-based data via RNN | AD & NC |
| Image Segments | Clas. | [2] | Multilevel threshold segmentation with GWO | AD & NC (MRI) |
| | Clas. | [25] | Integrate voxel-based, region-based and patch-based data into the CNN model | AD & sMCI & pMCI & NC |
| Transfer Learning | Clas. | [27] | Image entropy and VGG pre-trained model | AD & NC & MCI (MRI) |
| | Clas. | [28] | Pretrained CNN model with LSTM trained with MRI on three planes | AD & NC |
| | Clas. | [29] | Pretrained custom CNN model | AD & NC |
| | Clas. | [30] | Pretrained VGG16 and Inception V4 | AD & NC |
| Others | C&R | [31], [32], [33] | Machine learning with hand-crafted features | |

Note AIT (Artificial Intelligence Technology), C&R (Classification and Regression), Reg. (Regression), Clas. (Classification), GWO (Grey Wolf Optimization), pMCI (progressive Mild Cognitive Impairment), sMCI (stable Cognitive Impairment), eMCI (early Mild Cognitive Impairment), lMCI (late Mild Cognitive Impairment) such as 2D and 3D MRIs on different planes and clinical scores to build multi-channel CNN to increase the model prediction ability [10] [11][13][16][25].

GCN is to apply CNN to the graph network data in the non-Euclidean space. Due to the irregular space in graph data, it is challenging to extract graph patterns with regular CNN. Some researchers examined Cheb-Chevy GCN to classify the entire graph network generated from the MRI or PEI [21] [41]. Different factors, such as imbalanced datasets or model architectures or incomplete data, could affect all models' performance. Therefore, the ensemble learning method can combine multiple models into one superior performance model. The researches with this technique focused on model training with different aspects of MRIs and ensembling techniques to boost the model performance [4] [7][23][24].

Early AD prediction, such as early Mild Cognitive Impairment (eMCI), can be related to patients' conditions such as age, gender, previous clinical score [5]. Thus, RNN interprets the time-series data or generates a dense representation of patients' demographic information. Besides, some specific brain regions are positively related to AD status, such as GM, WM, Hippocampus; these regions' anatomical changes could reflect AD more effectively than other regions. Some researchers applied image segmentation algorithms to highlight MRI's sensitive brain region for further accurate prediction [2][26]. Transfer learning adapts a pre-trained model to new training data, and improve the training speed and model performance [27][28][29][30]. Table II shows the contributions of the reviewed studies.

### B. Preprocessing Techniques of Medical Imaging Dataset

Raw MRI or PET has no outstanding image quality due to various radiations and imaging equipment characteristics. Scanned images of AD patients show various changes in different sub-regions of the brain. Based on the previous researches, certain sensitive regions can reflect the AD symptoms more accurately, such as Hippocampal Regions (HR), Grey Matter (GM), White Matter (WM) and Cerebrospinal Fluid (CSF).

Therefore, most of the latest research process MRIs or PETs with comprehensive techniques when predicting AD's early stage via AI technologies. In this review, there are three categories of data preprocessing techniques employed before proposed methods were implemented, including Region Of Interests (ROI), Landmark Patches Extraction (LPE), and Regular Preprocessing (RP).

ROI's purpose is to segment the most critical brain regions related to AD on MRIs or PETs. One type of ROI is to extract the exact volume of certain regions such as HR in 3D and then slice the volume into 2D images. The other type is to acquire 2D images at first and then segment regions such as GM, WM, CSF on images with the medical imaging software or the MATLAB toolbox. The segmentation of sub-regions can highlight features of images to improve the interpretability of the AI models. Except for crucial brain regions such as GM, WM, CSF, HP, there are other tissues reflecting AD status at a certain level, and it is challenging to segment all of them via ROI techniques. To capture essential features, LPE puts landmarks on various sensitive regions of MRIs via specific algorithms, and small size patches of landmarks are extracted to form the dataset for testing and training from MRIs.

Both ROI and LPE techniques need a sturdy hypothesis of the brain regions and professionals post-check of data segmentation. If either fails to meet data quality control criteria, model performance cannot be reliable in the real application. It is also challenging to have a large amount of processed MRI images checked by certified professionals. All RP techniques are applied before ROI and LPE in all researches. However, RP techniques are more practical when AI technologies are applied to a large amount of data, and it is easier to integrate the preprocessing into the data pipeline of proposed methods. More details on the data preprocessing techniques categories are shown in Table III.

## C. Limitations and Future Direction

Traditional machine learning algorithms had been widely studied with hand-crafted features to predict AD previously. In contrast, the latest AI technology, such as CNN, RNN, can achieve an even better performance via high-level automatic feature extraction of the dataset. The advantage of the deep-learning technology is that it can either combine hand-crafted features to feature maps of input data or extract specific input data patterns directly for both classification or regression tasks. The AI technology can be integrated into a high-level application without specific features extracted via certified professionals from the data source.

However, there are certain limitations when applying deep-learning technologies in AD patients' classification or regression as follows:

1) CNN maps MRIs to their labels via updating weights of kernel filters during the convolutional computation, and it can extract common patterns of data with the same labels very well. However, it cannot learn the relationship among the individual input data and affect the prediction based on this relationship.

TABLE III
CATEGORIES OF MRI OR PET PREPROCESSING TECHNIQUES

| Type | Data Sources | Ref. | MRI or PET Preprocessing |
|------|--------------|------|--------------------------|
| ROI | ANDI, AIBL | [3] | Segment bilateral HR in MNI via 'Local learning algorithm' and extract regions in 3D bounding box; |
| | | [9] [15] [16] [12] | Registration with MNI. Normalization with DARTEL. Segmentation with SPM toolbox; |
| | | [22] [29] [2] | Register to MNI space and use AAL2 atlas consisting of ROIs for PETs. Skull stripping, contrast enhancement, histogram equalization and segmentation of GM, WM, CSF with SPM for MRI; |
| | | [25] [4] | Perform 'Anterior Commissure (AC) – Posterior Commissure (PC)' followed by 'Skull Stripping' and 'Cerebellum Removal.' And align each image to the 'Colin27 template' and use the 'N3' algorithm for intensity inhomogeneity correction; |
| | | | Segment GM, WM and CSF with FAST and divide each segmented image into regions via HAMMER[8]; |
| LPE | ADNI, MIRIAD, AIBL | [13] [11] | Perform 'Anterior Commissure (AC) – Posterior Commissure (PC)' followed by 'Skull Stripping' and 'Cerebellum Removal.' Moreover, align each image to the 'Colin27 template' and use the 'N3' algorithm for intensity inhomogeneity correction. Then extract informative image patches of multiple anatomical landmarks via the 'Data-Driven Landmark Detection Algorithm. |
| RP | ANDI, OASIS | [8] | Use 'Histogram Equalization' to generate a contrast-adjusted of original MRI and then use the Threshold Selection Method' from a grey-level histogram to strengthen the contrast of MRI |
| | | [10] | Select five axial slices for each subject individually for training directly |
| | | [14] | Pixel normalization and image augmentation with certain angles |
| | | [21] | The process is done via GRETNA[9]. Spatial normalization via DARTEL |
| | | [23] [27] | Image entropy to pick informative images. |
| | | [24] | Normalize the image data with SPM following a template of the 'International Consortium for Brain Mapping.' |
| | | [7] | Both MRI and PET are smoothed, averaged, spatially aligned, AC-PC orient baseline corrected, interpolated to standard voxel size and intensity normalized |
| | | [28] | Intensity Normalization (zero-center) and registration to MNI using SPM12 toolbox |

2) One of the challenges in AD prediction research is the lack of training data. Although research uses ADNI data in this review, each categorical data is limited, and AI model performance relies on a large amount of high-quality data for better performance.

3) In this review, all proposed methods involved CNN never discussed balancing the dataset of each label. Imbalanced datasets can bias the prediction to label what the model sees most during training, which gradually undermines model performance. It is also challenging to maintain a balanced dataset when the model is implemented in practice.

---

[8] https://www.med.unc.edu/bric/ideagroup/free-softwares/fast-hammer/
[9] https://www.nitrc.org/projects/gretna/

4) The ensemble learning method can boost model performance while it is challenging to apply it in practice. The transfer learning technique can improve both training speed and performance, but datasets' size restricts it. This limit might cause overfitting, and the structure of the pre-trained model might not perform best.

5) Most of the reviewed methods focus on supervised learning where the MRIs or PETs missing labels are discarded. Some researchers apply unsupervised learning techniques to compress and cluster the dataset. However, low-dimension data cannot interpret each other's relations, and the lack of data embeddings decreases the clustering performance.

6) MRIs or PETs preprocessing can be very complicated and computationally expensive. The precise segmentation also needs qualified checkout via professionals, which is not sustainable for a large dataset. The segmenting process needs third-party software or toolbox, challenging to integrate into the framework's data pipeline.

Future research needs to focus on applying semi-supervised learning algorithms to take advantage of labelled and unlabeled datasets. The dataset should also be processed with conventional processing techniques. The graph data or image embeddings need to be generated to represent relationships that can link supervised and unsupervised learnings. Finally, model training should take account of both labelled data and the relationship to this data so that a small dataset or imbalanced dataset has the lowest effect on the model's performance.